# Dynamically suppressed lattice rotations in $SrTiO_3$ as a basis for photo-induced ferroelectricity

Huaiyu Hugo Wang*[1,2], Michael Fechner*[3], Giovanni De Vecchi[3], Sylvia L. Griffitt[4], Gal Orenstein[1,2], Jade Stanton[1,2], Viktor Krapivin[1,2], Man T. Wong[5], Zhuquan Zhang[5], Mina Bionta[1], Vincent Esposito[1], Meredith Henstridge[1], Matthias C. Hoffmann[1], Patrick L. Kramer[1], Zach Porter[1], Ryan A. Duncan[2], Takahiro Sato[1,2], Soyeun K. Kim[1,2], Hasan Yavas[1,2], Samuel Teitelbaum[6], Keith Nelson[5], Ankit S. Disa[4], Michael Först[3], Mariano Trigo[1,2], Andrea Cavalleri[3,7]

* These authors contributed equally to this work.

*[1]Linac Coherent Light Source, SLAC National Accelerator Laboratory, Menlo Park, CA, USA*
*[2]Stanford Pulse Institute, SLAC National Accelerator Laboratory, Menlo Park, CA, USA*
*[3]Max Planck Institute for the Structure and Dynamics of Matter, Hamburg, Germany*
*[4]School of Applied & Engineering Physics, Cornell University, Ithaca, NY USA*
*[5]Department of Chemistry, Massachusetts Institute of Technology, Cambridge, MA USA*
*[6]Applied Structural Discovery, Arizona State University Biodesign Institute, Arizona, USA*
*[7]Department of Physics, Clarendon Laboratory, University of Oxford, Oxford, UK*

Photo-induced ferroelectricity in the quantum paraelectric $SrTiO_3$ involves the dynamical interplay between a coherently driven Ti–O stretching vibration and multiple structural degrees of freedom, including antiferrodistortive rotations, strain, and the polar mode instability. In the high-temperature cubic phase, in the absence of average antiferrodistortion, time-resolved X-ray diffuse scattering has evidenced a correlation between a photo-induced reduction in antiferrodistortive fluctuations and the emergence of ferroelectric order. Here, we complement these measurements with time-resolved elastic X-ray diffraction in the low-temperature tetragonal phase, in which antiferrodistortive fluctuations are small but a finite average rotation has set in. In this phase, we observe a long-lived reduction of the equilibrium antiferrodistortive rotation angle. A unified theory of the nonlinear lattice dynamics based on first-principles calculations describes the dynamics in both high-temperature cubic and low-temperature tetragonal phases, providing a basis for light-induced ferroelectricity in $SrTiO_3$.

Cubic perovskites frequently exhibit multiple instabilities toward more than one ground state, manifested in a competition between polar (ferroelectric) and antiferrodistortive phases. In $SrTiO_3$, the high-temperature cubic phase hosts a soft polar mode and is close to a ferroelectric transition, but does not develop long-range polar order at any temperature. Rather, at 110 K the crystal symmetry is lowered to a tetragonal phase, characterized by antiferrodistortive (AFD) rotations of the $TiO_6$ octahedra and unit-cell doubling. This effect likely contributes to shifting the mean field ferroelectric transition to lower temperatures ($T \approx 40$ K) [1,2], where polar fluctuations emerge and further suppress ferroelectricity.

When approaching the 110 K tetragonal transition from above, softening of the antiferro-distortive mode – located at the Brillouin-zone boundary – leads to a strong enhancement of rotational fluctuations. Below the transition, these fluctuations are rapidly suppressed and long-range rotational order develops, as shown in Fig. 1a. The susceptibility of $SrTiO_3$ to become ferroelectric is in fact suppressed by both the high-temperature rotational fluctuations and the low-temperature rotational order [1]. To qualitatively illustrate this interplay, we compute the temperature-dependent ferroelectric soft-mode potential using frozen-phonon calculations within density functional theory (DFT), integrating out the rotational degree of freedom including its fluctuations. The resulting potentials are shown in Fig. 1b, evidencing that AFD order and fluctuations reduce the depth of the double-well potential associated with the ferroelectric instability. Small perturbations modify the delicate balance between these competing lattice instabilities. Indeed, isotope substitution [3], chemical doping [4], or epitaxial strain [5] are sufficient to stabilize ferroelectric order in $SrTiO_3$.

Selective optical excitation of phonons [6,7] is an additional mechanism that perturbs this balance, in analogy to many other manifestations of induced order in superconductors, magnetic and ferroic materials [6–18]. In the case of $SrTiO_3$, it was shown that coherent excitation of a high-frequency Ti-O stretching vibration near 20 THz causes the emergence of a metastable ferroelectric phase [6], evidenced for example by the appearance of a second-harmonic generation (SHG) from near-infrared probe pulses.

The base temperature dependence of the induced SHG intensity is shown in Fig. 2. For all temperatures between 3 K and room temperature, it reduces continuously. In this temperature dependence, it is surprising that neither the tetragonal transition at 110 K nor the cross-over to quantum paraelectric behavior at 40 K are visible, showing that one cannot understand the phenomenon of photo-induced $SrTiO_3$ ferroelectricity by considering the perturbation of either the equilibrium antiferrodistortive order or the quantum polar fluctuations, both of which strongly evolve across these temperature ranges.

Here, we combine the same mid-infrared excitation that induces the ferroelectric SHG signal with time-resolved X-ray scattering, to probe the dynamics of the absolute AFD rotation angles in the low-temperature tetragonal phase (T<110 K, see Fig. 3a,c) and correlate them with photo-induced ferroelectricity. The experiments were performed at the X-ray pump-probe beamline of the Linac Coherent Light Source [19–21] [19] [19], with the sample mounted inside a liquid helium cooled cryogenic vacuum chamber. Linearly polarized, monochromatized X-ray probe pulses of 12 keV photon energy were combined with mid-infrared (mid-IR) excitation pulses of 22 THz central frequency and fluences up to 120 $mJ/cm^2$. The excitation pulses were tuned to drive the highest-frequency infrared-active Ti-O stretch phonon, extending to the Reststrahlen band of the 17-THz transverse-optical mode frequency, and their polarization was parallel to the [001] crystallographic direction. To match the photo-excited and probed volumes while maximizing the excitation fluence, we used a grazing-exit geometry with an exit angle set to 2° with respect to the sample surface.

Fig. 3c shows the time-dependent integrated scattering intensity at the (-3.5, -2.5, 4.5) Bragg reflection (cubic notation), which is sensitive to the amplitude of the AFD rotation, measured at a representative temperature of 75 K, well below the tetragonal transition. The resonant phonon excitation induces a prompt increase in the scattering intensity, accompanied by a few-cycle oscillation and followed by a long-lived reduction in intensity. The transient increase in scattering intensity corresponds to an enhancement of the AFD rotation angle, whereas the subsequent decrease reflects a persistent reduction, estimated to be approximately 0.3° at this temperature based on the calibration to the static X-ray scattering related to structural data taken from Ref. [22] as discussed in the supplement.

In a previous report [23], time-resolved X-ray diffuse scattering measurements taken in the cubic phase under the same mid-IR excitation were used to correlate the dynamics of the AFD fluctuations with photo-induced ferroelectricity. Figure 3d summarizes these measurements. The integrated diffuse scattering intensity at the R point—corresponding to the Brillouin-zone boundary location of the AFD soft mode—reacts to the excitation of the Ti–O stretch phonon. Similar to what is observed in the low-temperature tetragonal phase, a prompt increase in scattering intensity was observed after optical excitation, followed by a reduction at longer delay times. Because the R-point diffuse scattering intensity reflects the thermal population of the AFD soft mode, its reduction corresponds to a suppression of antiferrodistortive fluctuations. Relating the detected changes in diffuse-scattering intensity to the corresponding changes in antiferrodistortive fluctuations indicates that, at sufficiently long time delays, these fluctuations decrease by approximately 10%.

Taken together, the long-lived response observed in low-temperature elastic X-ray scattering reported here (Fig. 3c) and in high-temperature diffuse X-ray scattering reported earlier (Fig.

3d) indicates that mid-infrared phonon excitation reduces either the antiferrodistortive order parameter or its fluctuations.

We explain this behavior by a single theoretical model of the nonlinear lattice dynamics following Ti-O mode excitation in the cubic and tetragonal phases. As discussed in Ref. [23] for the high-temperature phase, excitation of the Ti-O stretching mode $Q_{IR}$ activates more than one nonlinear response. First, the resonantly driven mode $Q_{IR}$ couples quadratically to long-wavelength acoustic modes, thereby generating an effective strain coordinate $Q_\eta$ via the interaction potential scaling as $\alpha Q_\eta Q_{IR}^2$ [6]. Second, the driven mode couples to AFD fluctuations $\sigma_\phi^2$. This interaction is biquadratic in the interaction potential and proportional to $\beta \sigma_\phi^2 Q_{IR}^2$. It parametrically amplifies AFD fluctuations at opposite wavevectors $\pm q$, thereby enhancing $\sigma_\phi^2$ fluctuations without producing a static rotational distortion $\phi_0$. Our first-principles calculations show that $\beta < 0$, implying that the finite $Q_{IR}$ oscillations soften the AFD mode and transiently enhance rotational fluctuations (Fig. 4b, upper panel). Third, the strain field generated at the Brillouin-zone center couples to the AFD mode via a linear–quadratic interaction potential of the form $\kappa\, Q_\eta \sigma_\phi^2$. In contrast to the softening of the AFD fluctuations by its direct coupling to the driven mode, the positive coefficient of this term hardens the AFD mode and suppresses its fluctuations (Fig. 4b, lower panel). Because the effective strain relaxes on the slow acoustic time scale, this interaction dominates the lattice response at long time delays.

From these three interactions, we deduce the general form of the fluctuation dynamics, schematically illustrated in Fig. 4d. At negative times, before optical excitation of the Ti-O stretch mode, the crystal lattice exhibits thermal equilibrium AFD fluctuations, represented by the red shaded area in the figure. Immediately after excitation, the direct coupling $\sigma_\phi^2 Q_{IR}^2$ enhances the AFD fluctuations, expanding the shaded area. At later times, the strain-mediated interaction $\kappa\, Q_\eta \sigma_\phi^2$ (with $\kappa > 0$) dominates, suppresses the AFD fluctuations, and pushes them below the equilibrium value. This behavior is consistent with the experimental diffuse X-ray scattering data above the tetragonal phase transition (Fig. 3d).

We next extend this model to the low-temperature tetragonal phase of $SrTiO_3$. Here, the fluctuating coordinate $\sigma_\phi$ is replaced by $\phi_0 + Q_\phi$, where $\phi_0$ is the equilibrium rotation angle and $Q_\phi$ represents its average modulation. This mode corresponds to the totally symmetric zone-center Raman-active $A_g$ mode, which has its origin in the condensation of the zone-boundary AFD soft mode at the unit-cell doubling tetragonal phase transition. Substituting this expression into the coupling terms for the tetragonal phase leads to additional contributions. Specifically, interaction with the driven Ti-O stretch phonon produces a cross term $\beta\, \phi_0 Q_\phi Q_{IR}^2$, which rectifies the coordinate $Q_\phi$ and transiently shifts the equilibrium angle. Figure 4a shows the resulting effect on the potential of the AFD distortion.

Similarly, substituting $\sigma_\phi = \phi_0 + Q_\phi$ into the strain–AFD coupling yields a three-phonon mixing interaction potential proportional to $\kappa\, \phi_0 Q_\eta Q_\phi$. Because the corresponding coupling constant has a positive sign, this interaction reduces the rotation angle, as illustrated by the AFD potential in Fig. 4a (lower panel). Higher-order potential terms lead to additional squeezing of the mode, similar to what is found in the cubic phase. Because the strain response is long-lived, the strain-mediated interaction dominates at long time delays and results in a persistent reduction of the rotation angle.

Figure 4c illustrates the resulting dynamics. The comparison between the cubic and tetragonal phases shows that translating the interaction terms between the modes into the respective symmetry-adapted coordinates underscores the similarity of the dynamics observed across the tetragonal phase transition. The key point is that the coupling between the different structural degrees of freedom is an intrinsic property of the crystal lattice and therefore in essence unchanged across the structural phase transition. As a result, both the fluctuations and the mode coordinates display qualitatively similar behavior above and below the transition, respectively.

Beyond this qualitative discussion, we quantify the dynamics described by this model by combining the previously reported solution for the cubic-phase AFD fluctuation dynamics $\sigma_{\phi}^{2}$ with the tetragonal-phase dynamics calculated here. In the tetragonal phase, we solve the coupled equations of motion for the driven mode $Q_{IR}$, the effective strain coordinate $Q_{\eta}$, and the rotational mode $Q_{\phi}$, treating these degrees of freedom as oscillators with corresponding eigenfrequencies and lifetimes. The eigenfrequencies are taken from experimental measurements, while the nonlinear coupling coefficients are obtained by combining the tetragonal first-principles calculations described in the Supplementary Information with the corresponding cubic-phase coupling coefficients reported in our previous work [23]. The lifetimes of the modes are treated as free parameters and were allowed to vary within a physically reasonable range. This range is chosen to be broader in the vicinity of the cubic–tetragonal phase transition, where anharmonic effects become more pronounced. The parameters used for the effective simulations, together with the constraints applied in the fitting procedure, are summarized in the Supplementary Information, where we also compare the calculated dynamics with the experimental results. Overall, the model quantitatively reproduces the observed dynamics across the entire temperature range.

The obtained dynamics allow us to reconstruct the ferroelectric (FE) soft-mode potential in the driven state over a wide temperature range. We recomputed the ferroelectric potential from first-principles calculations while integrating out the dynamic rotation coordinate. Figure 5 shows the results for three representative temperatures (i) below the quantum paraelectric crossover, (ii) below the tetragonal transition, and (iii) in the cubic phase. In all cases, we find that the reduction of the AFD rotation angle, or of the AFD fluctuations in the cubic phase, lowers the FE soft-mode potential into a deeper double-well potential, which favors ferroelectric ordering. The depth of this potential depends on temperature, and at very low temperatures the double-well depth may become small.

In summary, we have presented a comprehensive experimental and theoretical picture of the dynamical response of antiferrodistortive rotations (T<110 K) and fluctuations (T>110 K) after coherent excitation of the high-frequency Ti-O stretching mode in $SrTiO_3$. We find that the same excitation reduces the role of these modes in both structural phases and contributes to enhancing ferroelectricity. Importantly, the experimental results are captured by a theory which describes both regimes using the same set of microscopic parameters, supporting the validity of these conclusions. Whilst these dynamics provide a basis for the interesting phenomenon of photo-induced $SrTiO_3$ ferroelectricity, the current picture does not consider the role of complex strain dynamics beyond our model and of a possible dynamical flexo-electric coupling, which is most likely to participate in these dynamics [6,24].

## Figures:

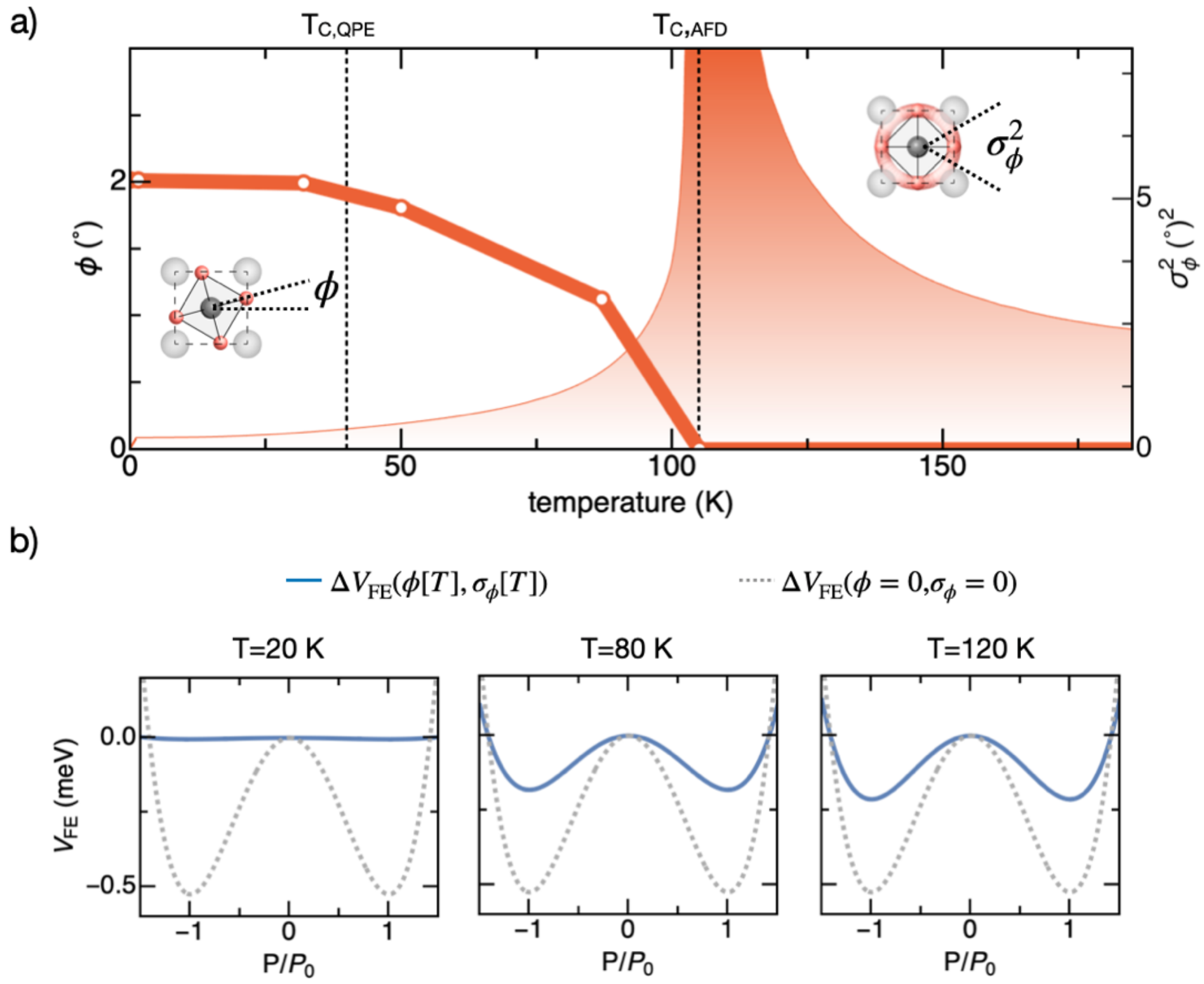

*Fig. 1. (a) Temperature dependence of the antiferrodistortive (AFD) order parameter (solid line) and its fluctuations (shaded area), as extracted from Refs.* [25,22]. *The softening of the AFD mode that drives the cubic-to-tetragonal transition is accompanied by an increase in fluctuations of the rotational order above $T_C$. Upon the onset of long-range rotational order and further cooling, these fluctuations are suppressed and nearly vanish at low temperatures. (b) Ferroelectric soft-mode potential at different temperatures (blue line), obtained from first-principles calculations by integrating out the rotational degree of freedom and its fluctuations as shown above. The gray dashed curve shows the bare ferroelectric potential in the absence of rotational order and fluctuations.*

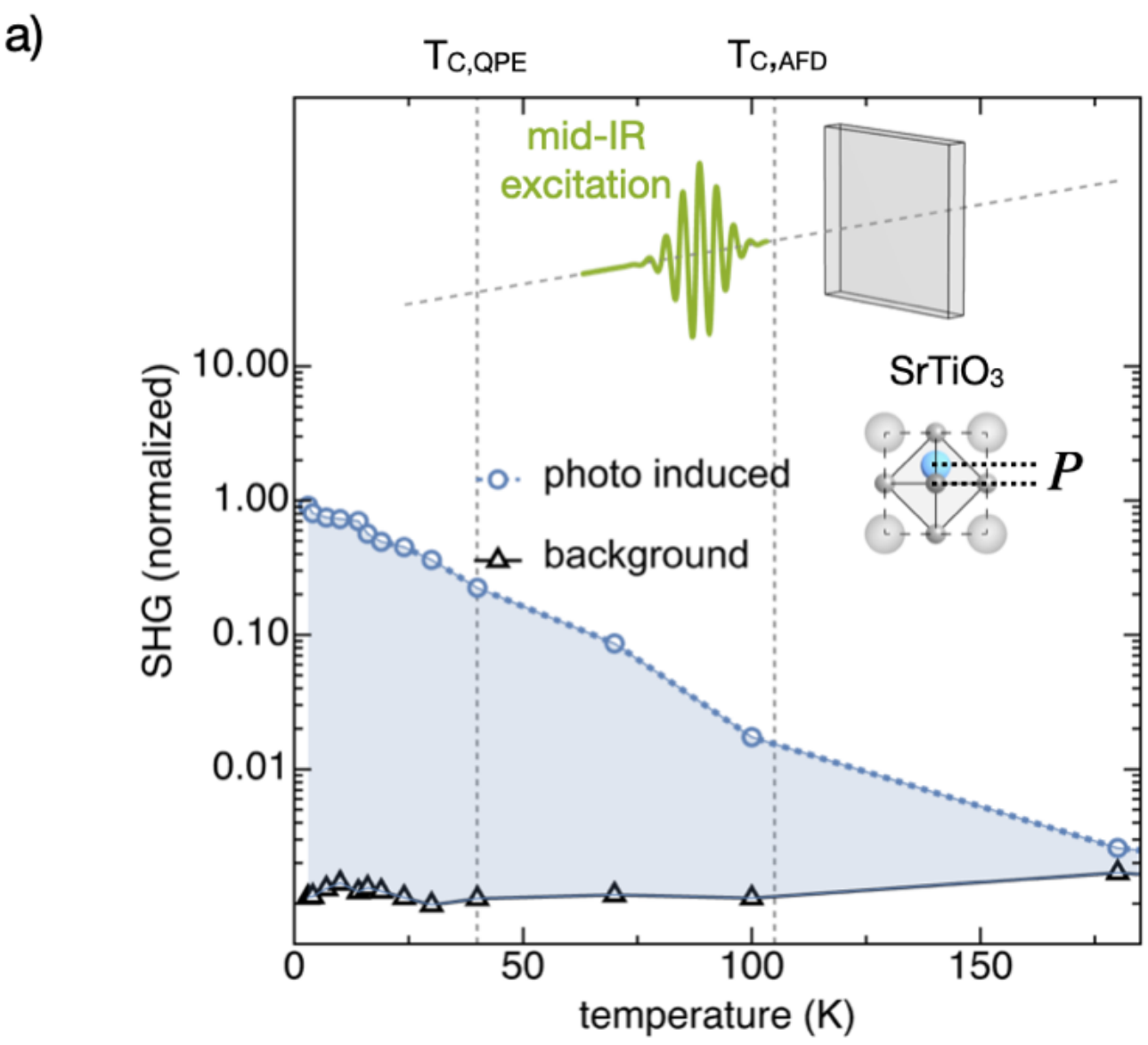

*Fig. 2. (a) Mid-infrared excitation of $SrTiO_3$ induces a metastable ferroelectric state. Temperature dependence of the light-induced state, measured by second-harmonic generation (SHG) taken from Ref.* [6]. *The light-induced state can be generated from cryogenic temperatures up to room temperature. Notably, the equilibrium transitions of $SrTiO_3$, indicated by the vertical dashed lines—including the quantum-to-classical paraelectric crossover and the tetragonal–cubic structural transition—have no discernible impact on the formation of the light-induced state*

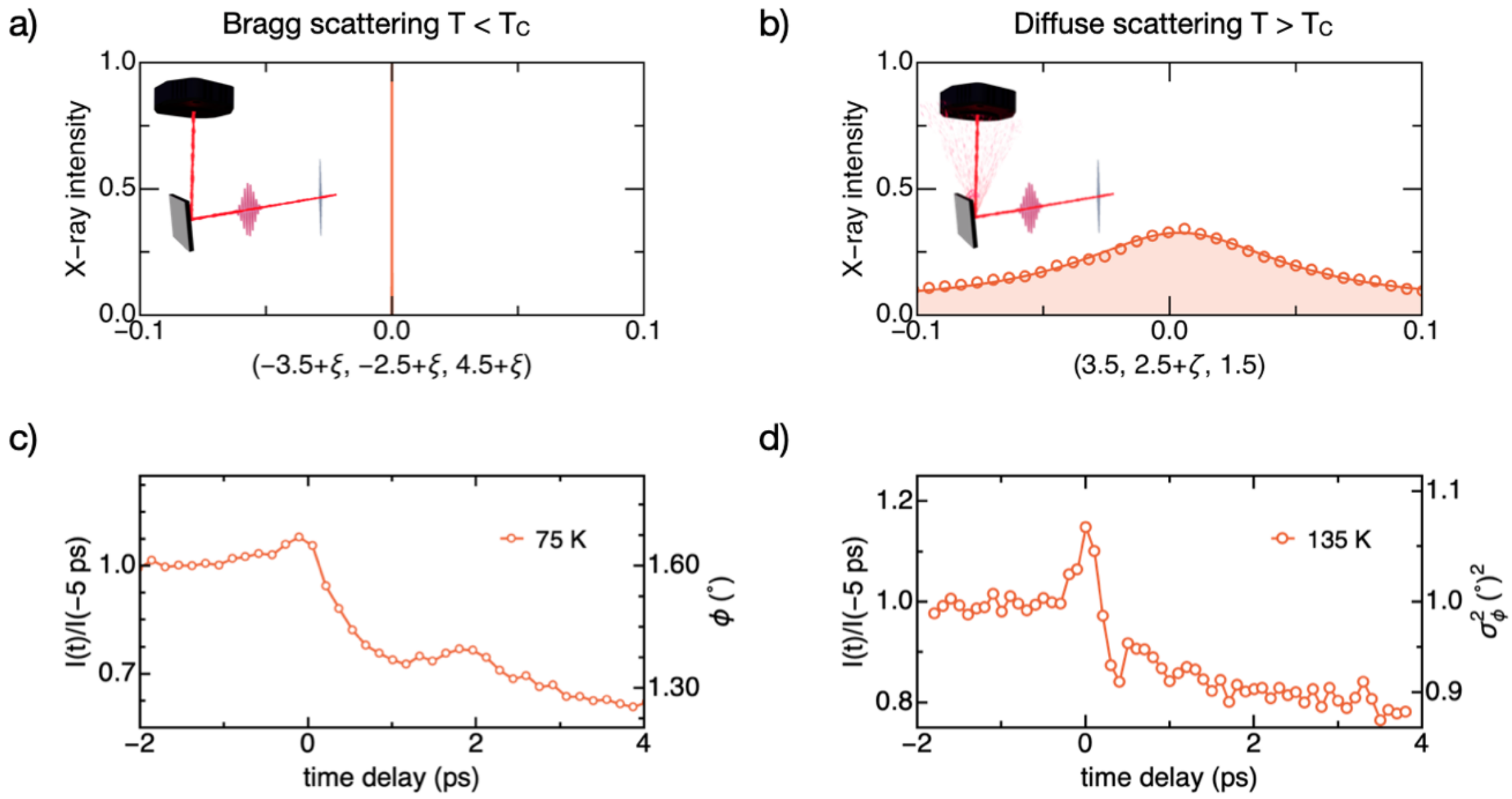

*Fig. 3. Mid-IR pump–X-ray probe experiments on $SrTiO_3$ above and below the antiferrodistortive (AFD) transition. (a) Below the transition, Bragg scattering at the cubic (-3.5,-2.5,4.5) peak directly probes the rotational structural order. (c) Time-resolved changes of the rotational peak induced by mid-IR excitation. Time zero marks the arrival of the mid-IR pulse. The left axis shows the change in integrated peak intensity, while the right axis shows the corresponding change in rotation angle extracted from the data. In addition to a small initial increase in signal corresponding to enhanced rotation, mid-IR driving induces a pronounced reduction of the rotational order by approximately 0.3°. (b) In the cubic phase above the transition, X-rays scatter from fluctuations at the (3.5,2.5,1.5) peak, allowing the magnitude of rotational-order fluctuations to be determined (data taken from Ref.* [23]*). (d) Pump–probe response of the diffuse scattering associated with rotational fluctuations. As in panel (c), the left axis shows the change in integrated diffuse intensity and the right axis the corresponding change in fluctuation amplitude (see Supplementary Material for details of the extraction procedure). Following an initial increase in intensity associated with enhanced fluctuations, the signal subsequently decreases, indicating a reduction of rotational fluctuations.*

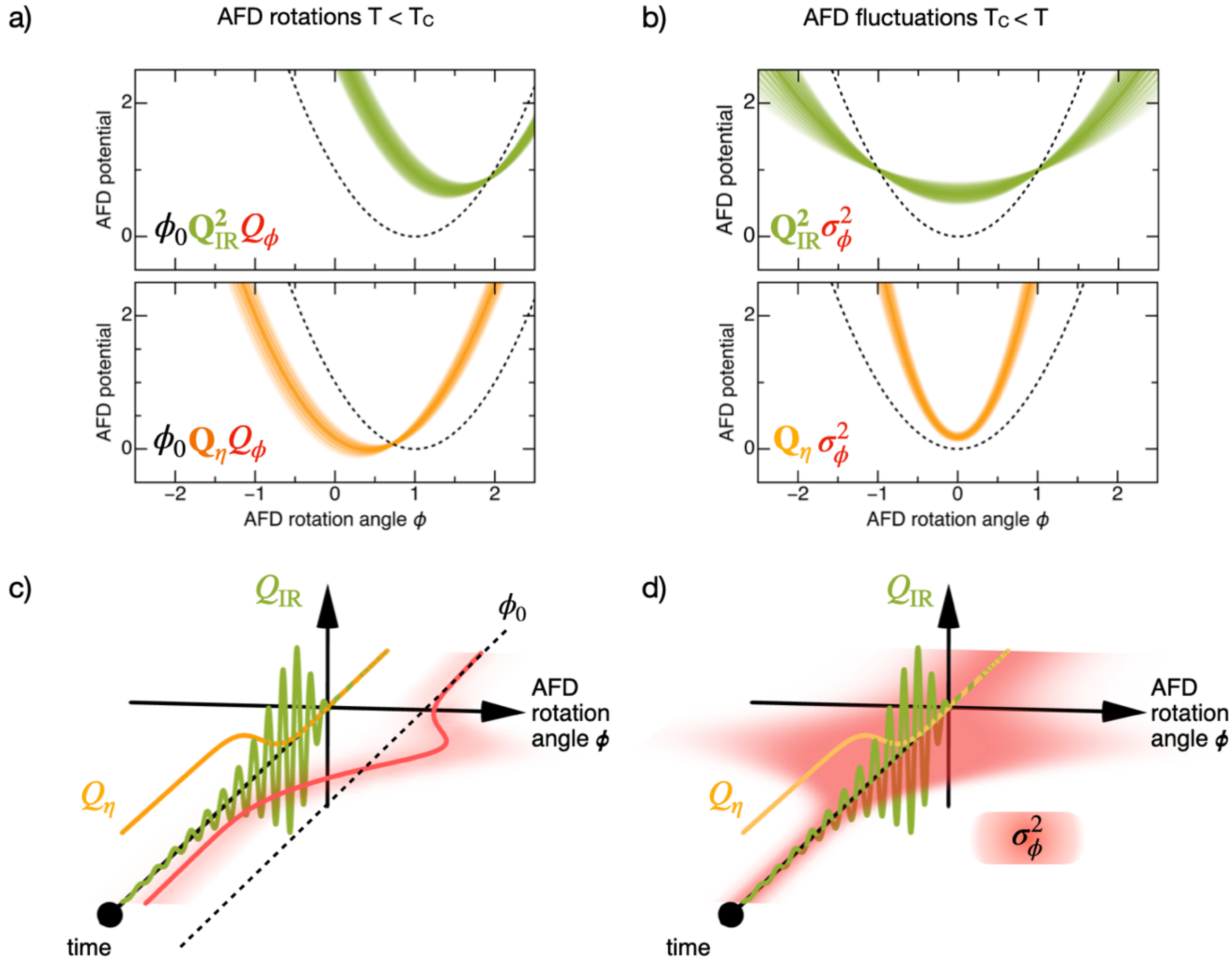

*Fig. 4: (a) Nonlinear coupling mechanisms between the driven mid-IR phonon (green) and the AFD rotation (red) and strain (orange) in the tetragonal phase. The phonon-induced displacement initially shifts the equilibrium rotation angle to larger values, while the strain response drives an opposite shift. Panel (c) shows the resulting time-dependent dynamics, where the fast relaxation of the optical mode and the slower strain response lead to a persistent long-term reduction of the rotation angle. (b) Nonlinear coupling mechanisms between the driven mid-IR phonon (green), AFD fluctuations (red shaded area in d), and strain (orange) above the structural transition. The optical-mode displacement increases the fluctuations, while the strain response suppresses them. Panel (d) shows the corresponding temporal evolution.*

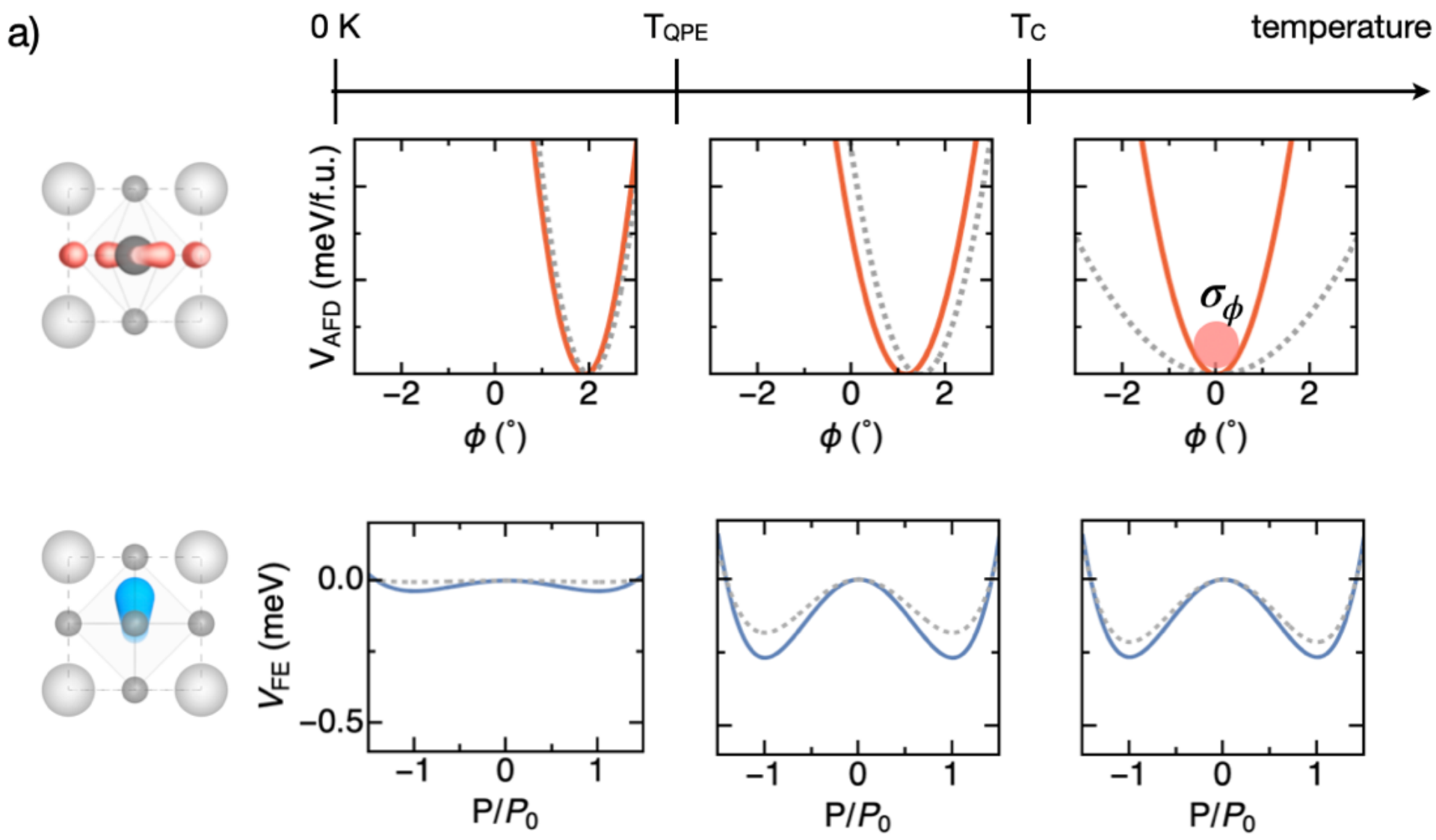

*Fig. 5: Mapped evolution of the AFD and ferroelectric potentials as a function of temperature. (a) Temperature dependence of the equilibrium (dashed line) and long-time AFD potentials (solid red line). (b) Equilibrium (dashed gray line) and long-time ferroelectric potentials (solid blue line) extracted from the same model, showing how the light-induced modification persists across the structural transition.*

# Supplementary information for

# Dynamically suppressed lattice rotations in $SrTiO_3$ as a basis for photo-induced ferroelectricity

Huaiyu Hugo Wang*[1,2], Michael Fechner*[3], Giovanni De Vecchi[3], Sylvia L. Griffitt[4], Gal Orenstein[1,2], Jade Stanton[1,2], Viktor Krapivin[1,2], Man T. Wong[5], Zhuquan Zhang[5], Mina Bionta[1], Vincent Esposito[1], Meredith Henstridge[1], Matthias C. Hoffmann[1], Patrick L. Kramer[1], Zach Porter[1], Ryan A. Duncan[2], Takahiro Sato[1,2], Soyeun K. Kim[1,2], Hasan Yavas[1,2], Samuel Teitelbaum[6], Keith Nelson[5], Ankit S. Disa[4], Michael Först[3], Mariano Trigo[1,2], Andrea Cavalleri[3,7]

* These authors contributed equally to this work.

*[1]Linac Coherent Light Source, SLAC National Accelerator Laboratory, Menlo Park, CA, USA*
*[2]Stanford Pulse Institute, SLAC National Accelerator Laboratory, Menlo Park, CA, USA*
*[3]Max Planck Institute for the Structure and Dynamics of Matter, Hamburg, Germany*
*[4]School of Applied & Engineering Physics, Cornell University, Ithaca, NY USA*
*[5]Department of Chemistry, Massachusetts Institute of Technology, Cambridge, MA USA*
*[6]Applied Structural Discovery, Arizona State University Biodesign Institute, Arizona, USA*
*[7]Department of Physics, Clarendon Laboratory, University of Oxford, Oxford, UK*

**Contents:**

S1: First-principles density functional theory calculations

We performed first-principles calculations within the framework of density functional theory (DFT) to explore the antiferrodistortive (AFD) and ferroelectric (FE) energy landscapes, the phonon excitation spectrum, the anharmonic coupling constants, and the phonon-induced stress of tetragonal $SrTiO_3$. All calculations were carried out using the Vienna *ab initio* Simulation Package (VASP 6.4) [1–3] in combination with the Phonopy software package [4,5] for phonon calculations. The electron–ion interaction was described using pseudopotentials generated within the projector augmented-wave (PAW) method [6]. Specifically, we used the standard Sr ($4s^2 4p^6 5s^2$), Ti ($3s^2 3p^6 3d^2 4s^2$), and O ($2s^2 2p^4$) PAW potentials. The exchange–correlation potential was treated within the revised Perdew–Burke–Ernzerhof generalized-gradient approximation for solids, PBEsol [7].

After convergence tests, we used a 12 × 12 × 12 Monkhorst–Pack k-point mesh for Brillouin-zone sampling and a plane-wave energy cutoff of 600 eV. The electronic self-consistency loop was converged until the change in total energy was below $10^{-8}$ eV. All calculations were based on the relaxed tetragonal structure of $SrTiO_3$, corresponding to the low-temperature antiferrodistortive phase with space group I4/mcm. The relaxed lattice constants were a = b = 5.49 Å and c = 7.82 Å, with Sr, Ti, O1, and O2 atoms occupying the Wyckoff positions b, c, a, and h(x = 0.725), respectively.

The phonon frequencies and eigenvectors were obtained from finite-displacement calculations using Phonopy. Non-analytical corrections were included in the dynamical matrix to account for long-range Coulomb interactions [8]. The phonon calculations were performed using a 2 × 2 × 2 supercell of tetragonal $SrTiO_3$. Since optical excitation couples directly to zone-center phonons, owing to the negligible momentum of photons on the scale of the Brillouin zone, we focused on the infrared-active polar modes at q = (0,0,0). To mimic the experimental conditions, we considered the highest-frequency polar mode of tetragonal $SrTiO_3$.

The anharmonic coupling constants were obtained using a frozen-phonon approach. To this end, we distorted a 2×2×2 supercell of tetragonal $SrTiO_3$ by superimposing selected combinations of phonon eigenvectors and homogeneous strain distortions. The resulting total-energy landscapes were fitted to polynomial expansions of the lattice potential. In particular, we extracted the nonlinear couplings involving the driven infrared-active mode $Q_{IR}$, the relevant AFD rotational coordinate $Q_{\phi}$, and the ferroelectric soft-mode coordinate $Q_P$. We also considered the coupling of strain, represented by the acoustic coordinate $Q_{\eta}$, to these structural degrees of freedom.

Finally, we mapped the ferroelectric energy landscape as a function of the AFD rotation amplitude. This calculation provides the dependence of the ferroelectric instability on the tetragonal rotation coordinate and was used to determine how the antiferrodistortive distortion modifies the ferroelectric polarization. Since the measurements were performed below the antiferrodistortive transition temperature, the relevant reference structure is the tetragonal phase with a finite equilibrium AFD rotation. The light-induced changes in the lattice potential are therefore described as modifications around this tetragonal ground state. We note that the DFT equilibrium rotation angle differs from the experimentally observed low-temperature value. We therefore mapped the DFT rotation coordinate onto the experimental rotation angle, using the calculated equilibrium angle of 6° and the measured low-temperature angle of 2° as reference points.

S2: Temperature dependence of photo-induced change in AFD rotations

The time delay dependent changes of the AFD rotation angles were extracted from elastic X-ray scattering measurements under the experimental conditions described in the main text for temperatures between 30 and 100 K. Figure S1 shows the absolute delay dependent scattering intensities, normalized to the absolute intensity measured at the lowest temperature. To reconstruct the rotational state of $SrTiO_3$, we calculated the temperature dependence of the (-3.5,-2.5,4.5) peak intensity as a function of the rotation angle using the lattice structure determination presented in Ref. [9]. Applying this calibration to our experimental time-resolved data allowed us to reconstruct the dynamics of the rotation angle shown on the right axes of Fig. 3c and Fig. S1.

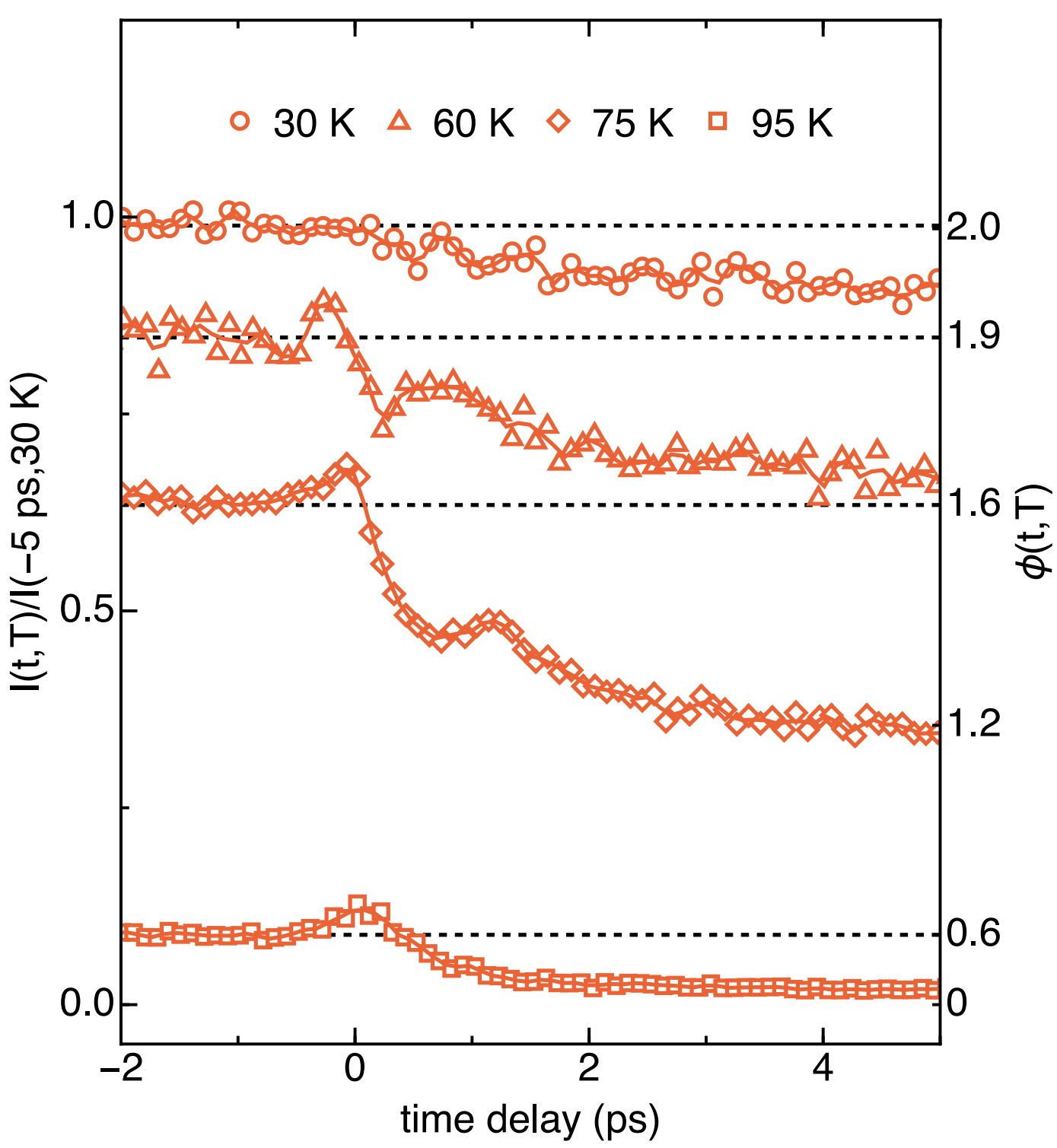

**FIG. S1: Time-resolved intensity changes of the rotational cubic (-3.5,-2.5,4.5) peak induced by the mid-IR excitation. The left axis shows the change in integrated peak scattering intensity normalized to the 30 K value, and the right axis shows the corresponding change in rotation angle extracted from a comparison of this data set to temperature dependent static data (see text) .**

## S3: Dynamics Simulations

As described in the main text, we used the following set of equations of motion to describe the dynamics of the rotational mode $Q_\phi$ in the driven state:

$$\frac{\partial^2}{\partial t^2} Q_{IR}[t] + 2\gamma_{IR} \frac{\partial}{\partial t} Q_{IR}[t] + \omega_{IR}^2 Q_{IR}[t] + \chi_{IR} Q_{IR}[t]^3 = Z^* E[t] \tag{S1}$$

$$\frac{\partial^2}{\partial t^2} Q_\eta[t] + 2\gamma_\eta \frac{\partial}{\partial t} Q_\eta[t] + \omega_\eta^2 Q_\eta[t] + \chi_\eta Q_\eta[t]^3 = \alpha\, Q_{IR}^2 \tag{S2}$$

$$\begin{aligned} \frac{\partial^2}{\partial t^2} Q_\phi[t] + 2\gamma_\phi \frac{\partial}{\partial t} Q_\phi[t] + \omega_\phi^2 Q_\phi[t] + \chi_{\phi_3} Q_\phi[t]^2 + \chi_{\phi_4} Q_\phi[t]^3 \\ = \beta \phi_0 Q_{IR}^2 + \kappa\, \phi_0 Q_\eta \end{aligned} \tag{S3}$$

The initial values for all phonon parameters were obtained from the first-principles calculations described in Section S1. The eigenfrequency of the $Q_\phi$ was taken from the experimental data reported in Ref. [10] , accounting for its temperature dependence. The initial value for the frequency of the strain coordinate $Q_\eta$ was estimated from the ratio between the penetration depth of the mid-infrared excitation pulses and the sound velocity [11], yielding a frequency in the GHz range. Hence, $Q_\eta$ represents a collective acoustic response rather than a single vibrational eigenmode, and consequently its eigenfrequency and lifetime should not be interpreted as those of an individual phonon mode. The same procedure was employed in our previous work [12].

Selected calculated phonon parameters, including damping coefficients, were then adjusted within physically reasonable boundaries to optimize agreement between the solutions of Eqs. (S1)–(S3) and the measured transient changes in X-ray scattering intensity. Specifically, we constrained the signs of the coupling coefficients to those obtained from DFT, restricted the strain frequency to the GHz range, and limited the maximum allowed strain to below 0.3%. Under these constraints, we obtained the fit shown in Fig. 3c and the temperature-dependent simulations shown in Fig. S2.

| $f_{IR}$<br>16 THz | $\gamma_{IR}$<br>1.6 THz | $\alpha$<br>$-3.1 \pm 0.2\ \mathrm{THz}^2 / (\sqrt{\mathrm{amu}}\mathrm{Å})$ |
|---|---|---|
| $f_\eta$<br>$(3 \pm 1)$ GHz | $\gamma_\eta$<br>$0.1 \pm 0.05$ THz | $\beta$<br>$(-9 \pm 0.5)\ THz^2 / (\mathrm{amu}\ \mathrm{Å}^2)$ |
| $f_\phi$<br>$taken\ from\ Ref.$ [10] | $\gamma_\phi$<br>$1 \pm 0.3$ THz | $\kappa$<br>$(12 \pm 5)\ \mathrm{THz}^2/(\sqrt{\mathrm{amu}}\mathrm{Å})$ |

**TABLE S1 | PARAMETERS OBTAINED FROM THE BEST MATCH OF THE SOLUTIONS OF EQUATIONS (S1)-(S3) TO THE EXPERIMENTAL DATA.**

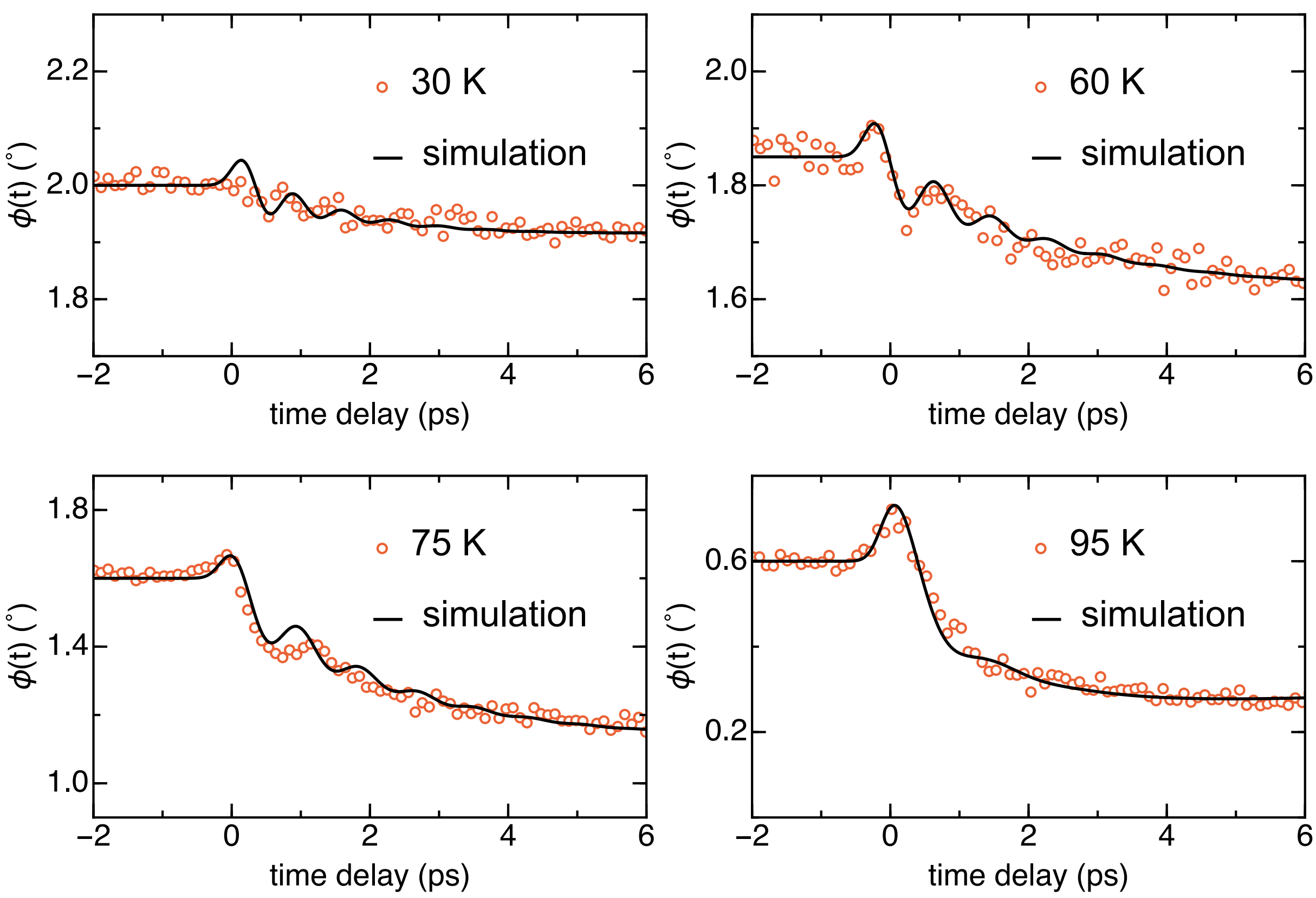

**Fig. S2: Comparison between the experimentally measured change in rotation angle and the numerical simulations obtained from Eqs. (S1)–(S3).**